\documentclass[aps,twocolumn,showpacs,preprintnumbers,amsmath,amssymb,prb,superscriptaddress]{revtex4-1}

\usepackage{amssymb}
\usepackage{graphicx}
\usepackage{amsmath}
\usepackage{color}
\usepackage{fancyhdr}
\usepackage{textcomp}
\usepackage{ulem}

\begin{document}

\title{Electrodynamics of Josephson junctions containing strong ferromagnets}

\author{D. Massarotti}
\email{massarottidavide@gmail.com}
\affiliation{Dipartimento di Ingegneria Elettrica e delle Tecnologie dell'Informazione, Universit\`{a} di Napoli Federico II, Via Claudio, I-80125 Napoli, Italy}
\affiliation{CNR-SPIN, Monte S. Angelo-Via Cintia, I-80126, Napoli, Italy}

\author{N. Banerjee}
\email{N.Banerjee@lboro.ac.uk}
\affiliation{Department of Physics, Loughborough University, Loughborough LE11 3TU, United Kingdom
}

\author{R. Caruso}
\affiliation{Dipartimento di Fisica "Ettore Pancini", Universit\`{a} di Napoli Federico II, Via Cintia, I-80126 Napoli, Italy}
\affiliation{CNR-SPIN, Monte S. Angelo-Via Cintia, I-80126, Napoli, Italy}

\author{G. Rotoli}
\affiliation{Dipartimento di Ingegneria Industriale e dell'Informazione, Universit\`{a} della Campania Luigi Vanvitelli, Via Roma, I-81031 Aversa (CE), Italy}

\author{M. G. Blamire}
\affiliation{Department of Materials Science and Metallurgy, University of Cambridge, 27 Charles Babbage Road, Cambridge CB3 0FS, United Kingdom
}

\author{F. Tafuri}
\affiliation{Dipartimento di Fisica "Ettore Pancini", Universit\`{a} di Napoli Federico II, Via Cintia, I-80126 Napoli, Italy}
\affiliation{CNR-SPIN, Monte S. Angelo-Via Cintia, I-80126, Napoli, Italy}

\begin{abstract}

Triplet supercurrents in multilayer ferromagnetic Josephson junctions with misaligned magnetization can penetrate thicker ferromagnetic barriers compared to the singlet component. Although the static properties of these junctions have been extensively studied, the dynamic characteristics remain largely unexplored. Here we report a comprehensive electrodynamic characterization of multilayer ferromagnetic Josephson junctions composed of Co and Ho. By measuring the temperature-dependent current-voltage characteristics and the switching current distributions down to 0.3 K, we show that phase dynamics of junctions with triplet supercurrents exhibits long (in terms of proximity) junction behavior and moderately damped dynamics with renormalized capacitance and resistance. This unconventional behavior possibly provides a different way to dynamically detect triplets. Our results show new theoretical models are required to fully understand the phase dynamics of triplet Josephson junctions for applications in superconducting spintronics.

\end{abstract}

\maketitle
  
\section{Introduction}

The competing nature of the superconducting and magnetic orders gives rise to a rich physics in superconductor (S)/ferromagnet (F) heterostuctures. Ferromagnetic Josephson junctions (JJs) are particularly interesting for their potential applications as switching elements in cryogenic memories \cite{birge2016,blamire2004,mukhanov,baek,newman,noi_JAP}, bi-stable states in quantum computation \cite{golubov,kawabata,kawabata2,feofanov,mqt} and circuit elements in superconducting spintronics \cite{efetov_rew,eschrig,birge2010,robinson,sprungmann,niladri,krasnov2014,aarts1}. Although the static properties of the SFS JJs have been extensively studied \cite{buzdin_rew,linder}, the dynamics of these junctions, especially those composed by strong ferromagnetic layers, remains to be explored.

SFS junctions with strong F layers like Co, can reach high ($\sim$ 100 $\mu$V \cite{robinson2006}) $I_c R_N$ values (where $I_c$ is the critical current and $R_N$ the normal state resistance of the junction) in the $\pi$ state \cite{ryazanov,buzdin_rew}, which can potentially be used for memory applications. Recently, JJs with multiple F layer barriers have been theoretically and experimentally studied in connection to unconventional triplet superconductivity with equal-spin Cooper pairs, that can be artificially generated in these structures \cite{robinson,niladri,linder}. The spin-aligned triplet Cooper pairs are immune to the exchange field of the F layer and the ability to transmit supercurrent through thick F layers has opened up the possibility to combine spin-based electronics with dissipationless superconductivity (superconducting spintronics \cite{eschrig,linder,birge2010,robinson,aarts1,aarts2}). However, the effects of multiple and complex barrier on the phase dynamics of the junction are completely unexplored. To the best of our knowledge, previous studies have only involved a single weak ferromagnet (PdNi \cite{aprili} or CuNi \cite{krasnov2007,goldobin1,goldobin2}), where the supercurrent transport is mediated by singlet Cooper pairs. Here, using a combination of the strong ferromagnet Co and spiral magnetic Ho layers, we have systematically studied fully metallic single (Co), bi (Ho/Co) and trilayer (Ho/Co/Ho) SFS JJ with Nb electrodes. Static properties of SFS junctions using Ho/Co/Ho composite barriers have extensively been studied in recent years as a model system exhibiting triplet superconductivity \cite{robinson,blamire2012}.

The dynamics of JJs is commonly understood in terms of the Resistively and Capacitively Shunted Junction (RCSJ) model \cite{barone,likharev}. In JJs where the weak link is an insulator, the large capacitance due to the dielectric barrier results in a hysteresis in the current-voltage (IV) characteristic of the junction. In the RCSJ framework, a SNS (N is a normal metal) junction should display an overdamped non-hysteretic behavior because of the negligible capacitance. However, there have been reports of hysteresis observed in SNS JJs attributed to heating in the normal part of the junction \cite{pekola}.

The energy scale which sets the critical current $I_c$ of an SNS junction is either the energy gap of the superconductor $\Delta$ or the Thouless energy\cite{thouless2,zaikin2001} $E_{th} = \hbar v_F l_e /3 L^{2}$. Here, $ \hbar$ is the reduced Planck constant, $v_F$ is the Fermi velocity, $l_e$ is the electron mean free path and $L$ is the separation between the superconducting electrodes. In mesoscopic transport terminology, short Josephson junctions are defined when $E_{th} > \Delta $, while $E_{th} < \Delta $ for long junctions \cite{thouless2,zaikin2001}. The Thouless energy also determines the minigap $E_g$ appearing in the density of states of the normal metal due to the proximity effect from the two superconducting electrodes \cite{golubov,skvortsov2015}.

In pure metallic SFS JJs, it is well known that the singlet pair correlations decay rapidly with increasing the thickness of the ferromagnetic layer, where the exchange field in the ferromagnet leads to phase decoherence of singlet Cooper pairs. Remarkably, for ferromagnetic barriers with multiple misaligned F layers, an equal-spin triplet component is induced with a decay length comparable to $\xi_N$ in the F layers \cite{efetov_rew}. These SFS junctions can, therefore, be treated as effective SNS junctions \cite{efetov_rew,golubovIV}.

We will show how a careful inspection of junction dynamics gives indications on the presence and on the effects of triplet currents in bilayer and trilayer junctions. Interestingly, these JJs show a hysteretic behavior in the IV characteristics at temperatures lower than 2 K, and high values of the $I_c R_N$ product up to about 500 $\mu$V. We have studied the dependence of $I_c$ and of the switching current distributions as a function of temperature over a wide temperature range. This comprehensive electrodynamic characterization and the comparative analysis provide evidence of a strikingly different behavior for bilayer and trilayer JJs compared to single layer junctions. The dynamics of the former class falls in the framework of long SNS JJs, which can possibly be explained as arising from the presence of long range triplet Cooper pairs in these junctions \cite{efetov_rew,golubovIV}.

The paper is organized as follows: in Section \ref{setup} the fabrication of SFS JJs containing Co and Ho and the measurement setup are presented. Section \ref{IV} describes the measurements of the IV characteristics while in Section \ref{energy} the temperature dependence of $I_c$ is analysed in the framework of the long junction regime. Finally, in Section \ref{phase} the measurements of switching current distributions are reported, which provide strong evidence of moderately damped dynamics and represent the key tool to reconstruct the electrodynamics of the junctions. Conclusions are summarized in Section \ref{disc}.
 
\section{Fabrication and measurement setup}
\label{setup}

The thin film stacks were grown in an ultra-high vacuum chamber using dc magnetron sputtering on unheated (001) Si substrates with a 250 nm thick SiO$_2$ coating. The base pressure of the chamber was maintained below 10$^{-7}$ Pa and the chamber walls cooled via a liquid nitrogen jacket. Before the actual deposition each target was pre-sputtered for 15-20 minutes to clean the surfaces. The entire stack was grown in a single run to ensure excellent interface quality. Devices were prepared using standard optical lithography and Ar-ion milling which were used to define 4 $\mu$m wide tracks. The tracks were narrowed down by focused-ion-beam milling to create current-perpendicular-to-plane devices: the details of the process are described elsewhere \cite{bell,niladrifab}. The actual device dimensions could be controllably varied by changing the width of the cuts but the average device dimensions were in the range of 500 nm to 600 nm. For all JJs, the lateral dimensions are less than half of the Josephson penetration depth and so magnetic self-field effects from the drive current can be neglected. The Co thickness in all the junctions was kept at 6 nm and the Ho thicknesses of for bi and trilayer junctions were fixed at 6 nm. The combined thicknesses of Co and Ho in bilayer and trilayer junctions ensure that the majority of singlet Cooper pairs are filtered out \cite{blamire2012,robinson2006}. In all junctions, the magnetic layers were separated from the 250 nm thick top and bottom Nb electrodes and from each other by 5 nm Cu layers (Fig. \ref{fig:1}a). The Cu layers ensure good magnetic properties of Co and Ho on Nb and magnetically decouple the F layers.

SFS junctions containing a F layer of thickness $L$ are in the dirty limit (diffusive) if the electron mean free path $l_e<L$ and $l_e< \xi_F$, where $\xi_F$ is the superconducting coherence length in the ferromagnetic layer \cite{blamire2012}. In this limit, the transport properties of the junctions can be understood in terms of the semiclassical diffusion-like Usadel equations, whereas in the ballistic (clean) limit, characterized by $l_e>L$ and $l_e>\xi_F$, the Eilenberger approach is more suitable \cite{blamire2012}. Bi and trilayer junctions containing 6 nm Ho and 6 nm Co are in the dirty limit since for Ho, $\xi_F$ $\simeq$ 4 nm and  $l_e$  $\simeq$ 0.87 nm \cite{blamire2012}. In this limit, the relevant energy scale is the Thouless energy as defined before.
It is not straightforward to analyze junctions containing only Co, since JJs containing 5 nm Co have shown clean limit behavior \cite{robinson2006}.

The transport measurements were carried out in a $^{3}$He Heliox cryostat with a base temperature of about 0.3 K. For filtering, a room temperature electromagnetic interference filter stage was used followed by low pass RC filters with a cut-off frequency of about 1 MHz anchored at 1.5 K, and by two stages of copper powder filters thermally anchored at the 1K-pot stage and at the sample stage \cite{prbLuigi}, respectively. Standard four-point resistance measurements as a function of temperature and current-voltage characteristics as a function of temperature and magnetic field have been performed. To gain a deeper insight into the electrodynamics of the junctions, we have performed the temperature-dependent measurements of the switching current distributions (SCDs). Here, the junction is current biased with a ramp at a constant sweep rate $\Delta I/\Delta t$, the voltage is measured using a low noise differential amplifier and is fed into a threshold detector, which is set to generate a pulse signal when the junction switches from the superconducting state to the finite voltage state \cite{prbLuigi}. This signal is used to trigger a fast voltmeter to record the value of the switching current. This procedure is repeated at least $10^4$ times at each temperature, which allows us to construct a histogram of the switching currents.

\section{Current-voltage characteristics}
\label{IV}

\begin{figure}[htb!]
\includegraphics[width=0.85 \linewidth]{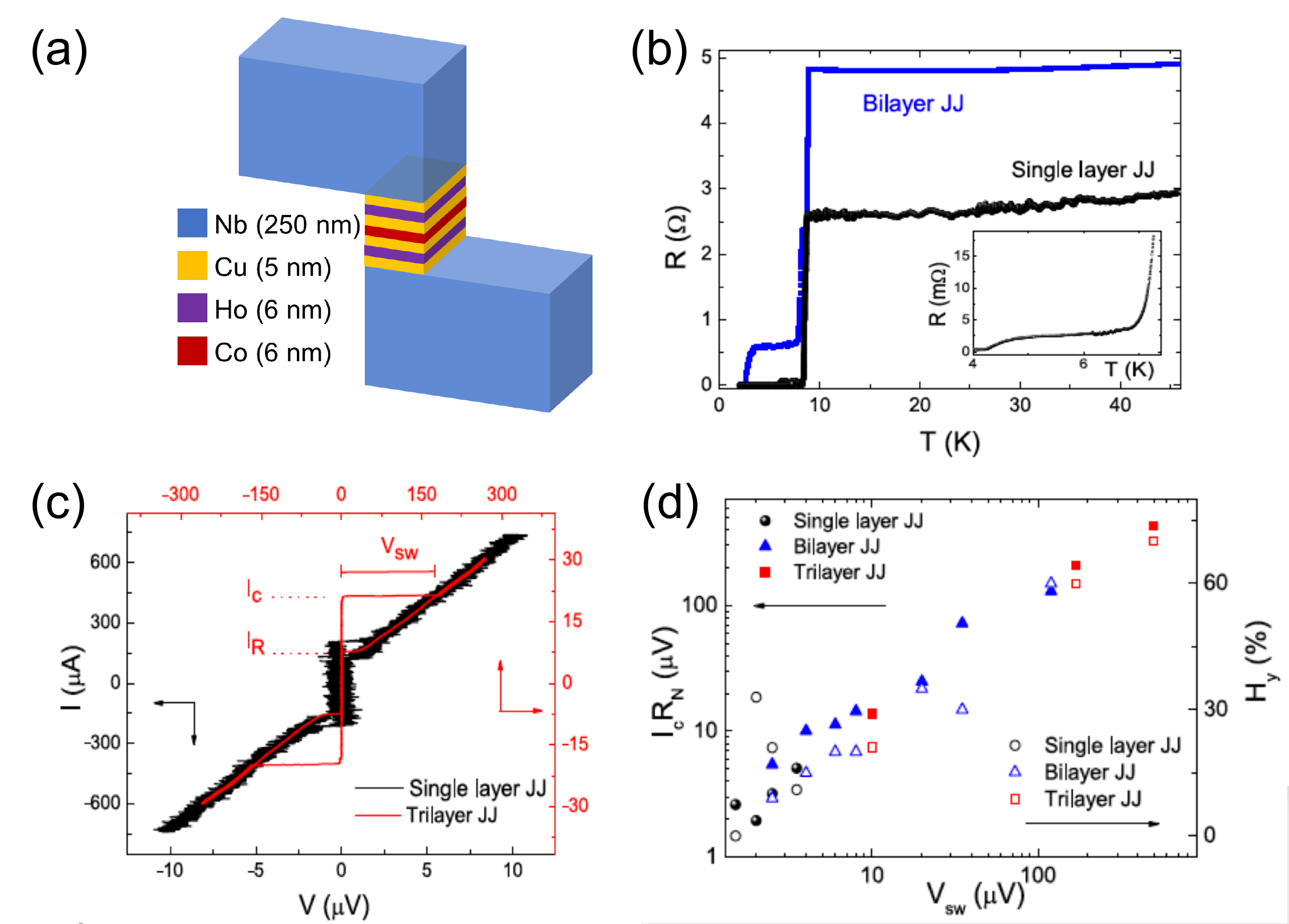}
\caption{a) Figure of a typical trilayer junction with the corresponding layer thicknesses. The single and bilayer junctions have similar Nb, Cu, Co and Ho thicknesses.  b) Resistance vs temperature behavior for single layer (black curve) and bilayer JJ (blue curve), respectively. The inset shows a magnified view of the single layer JJ curve, below 7 K, which highlights the superconducting transition at about 4 K. c) IV characteristics for single layer JJ (black line, left and down axis) and for trilayer JJ (red curve, right and top axis), respectively. Both IV curves have been measured at 0.3 K. Definition of $I_c$, $I_R$ and $V_{sw}$ are indicated on the IV curve of the trilayer JJ. Note that the two axes have different current and voltage scales. d) $I_c$$R_N$ product as a function of $V_{sw}$ (left axis) for single layer (black dots), bilayer (blue triangles) and trilayer (red squares) JJs, respectively. The hysteresis $H_y$ for the same junctions is shown by the corresponding open symbols (right axis).}
\label{fig:1}
\end{figure}

Junctions fabricated from stacks with various F layer combinations were measured down to 0.3 K. The transport properties of some representative junctions are summarized in Table \ref{tab:1}. In Figure \ref{fig:1}b, two $R$ vs $T$ curves for a single (black) and bilayer (blue) junction are shown. For both junctions, the electrodes become superconducting at about 8 K. This is followed by a region where the resistance has a few Kelvin wide tail down to the critical temperature ($T_c$) of the junction, which is usually below 4 K. For the single layer junction, a magnified view of the $R$ vs $T$ curve is shown in the inset of Fig. \ref{fig:1}b. Figure \ref{fig:1}c shows the IV characteristics of a single layer junction (black curve, left and down axis) and a trilayer junction (red curve, right and top axis), measured at 0.3 K. The switching voltage $V_{sw}$, defined as the voltage jump once the junction switches to the finite voltage state, and the normal state resistance $R_N$, measured as the linear slope of the IV curve above $V_{sw}$, both increase with the number of F layers. In Table \ref{tab:1}, the transport parameters, such as $I_c$, $J_c$ and the $I_c R_N$ product, are reported for some of the measured junctions.

We note that the values of the $I_c R_N$ product for bilayer and trilayer junctions are significantly higher than the estimated values for supercurrents composed purely of singlets. The singlet $I_c$ in a SFS junction has multiple oscillations as a function of the F layer thickness due to $0$-$\pi$ transition which is superimposed on an exponential decay envelope. To set an upper limit of the singlet $I_c$, we neglect the oscillations and take the exponentially decaying maximum $I_c$ envelope \cite{robinson2006}. This assumes that the decay in $I_c$ is limited by the coherence length and not by the dephasing of the singlet Cooper pairs \cite{niladri}. The singlet coherence length of Co and Ho is 3 nm and 4 nm respectively, which gives a total barrier thickness equivalent to 15 nm Co for the trilayer device. If $I_c$ in bilayer and trilayer junctions were dominated by singlets, for a 15 nm Co barrier we would expect a maximum value of the $I_c R_N$ product of about 0.4 $\mu$V. However, we observe a characteristic voltage at least 2-3 orders of magnitude higher for bilayer and trilayer devices, as reported in Table \ref{tab:1}. The high $I_c R_N$ values strongly indicate that the supercurrent is dominated by triplets \cite{robinson,birge2010}. While triplets have been widely reported in trilayer JJ, bilayer junctions have also been shown to generate triplets \cite{krasnov2014}. Theoretically, triplet generation in bilayer devices can arise from anomalous Andreev reflections, as recently reported in Refs. \onlinecite{houzet,scientific2016}.

Even for fully metallic JJs the IV characteristics strikingly show a finite hysteresis, quantified as $H_y=1- I_R / I_c $, where $I_R$ is the retrapping current. Hysteresis in the IV characteristics is routinely observed in SIS tunnel JJs (where I is an insulating layer) and is commonly described in a variety of physical conditions in terms of the RCSJ model  \cite{barone,likharev}: a large capacitance $C$ arising from the insulating barrier results in an underdamped dynamics and high values of the junction quality factor $ Q = \omega_p R C \gg 1$, where $\omega_p=(2eI_{c}/\hbar C)^{1/2}$ is the plasma frequency and $e$ the electron charge. However, in JJs with metallic weak links, the very low value of the geometric capacitance $C_g$ results in a quality factor $ Q  \ll 1$. This leads to an overdamped dynamics and the IV characteristic is expected to be non-hysteretic.

There have been reports of hysteresis in JJs with metallic weak links: for example, hysteresis in SNS junctions has been explained in terms of heating due to the high Joule power dissipated in the normal metal weak link, which raises its electron temperature \cite{pekola}. The increase of the electron temperature due to this heating causes a reduction of the retrapping current compared to its intrinsic value as the current is ramped down after switching, resulting in a finite hysteresis \cite{pekola}. An alternative explanation is provided in terms of the RCSJ model with a renormalized capacitance value. For specific junction configurations, the intrinsic capacitance $C_i$, introduced in Refs. \onlinecite{zaikin2010,skvortsov2015} to include the response of Andreev bound states to nonstationary boundary conditions beyond the tunnelling limit, may become the dominant capacitive term over $C_g$. This is especially relevant in SNS JJs where the geometric capacitance due to Coulomb interactions can be negligibly small. Therefore, although the junction resistance $R_N$ is about 1 $\Omega$ or less and $C_g$ is of the order of 1 fF, high values of the intrinsic capacitance $C_i$ provide an effective quality factor $Q$ higher than 1 \cite{zaikin2001,zaikin2010,skvortsov2015}.

Particularly for the long JJs relevant for this work, the strength of the proximity effect is characterized by the value of the spectral minigap  \cite{zaikin2010,skvortsov2015} $E_g \simeq \hbar / \tau_N $, where $\tau_N $ is the time required for an electron in the normal region to establish a contact with superconductors. This implies that the pair relaxation time in the weak link is essentially the diffusion time of the Andreev pairs in the normal region \cite{angers2008} and is given by $\tau_N = R C_i $ instead of $R C_g$.

To our knowledge, hysteresis in the IV curves of fully metallic SFS junction has been reported only in Ref. \onlinecite{krasnov2007} and the origin of such effect has been explained by considering the large overlap capacitance ($\simeq$ 35 pF) arising from the specific junction geometry, where the weak ferromagnetic CuNi layer acts as a ground plane for the JJ. An important point to note here is that most of the SFS junction measurements reported in literature have been carried out at 4.2 K, while the junctions measured in this work and in Ref. \onlinecite{krasnov2007} show a finite hysteresis only up to about 2 K (see Fig. \ref{thouless}b for a bilayer junction composed of Ho and Co). This implies that the presence of hysteresis in the IV curves results from the temperature dependence of the proximity effect. We highlight here that this hysteresis is different from that observed in other types of ferromagnetic-based JJs \cite{mqt,blamire2011,mukhanov,aprili,goldobin1}. In these cases the hysteresis could be clearly attributed to the insulating nature of the barrier, composed by a ferromagnetic insulator like GdN \cite{mqt,blamire2011} or by an insulating (Al/AlO$_x$) and a ferromagnetic metallic layer in SIFS junctions \cite{mukhanov,aprili,goldobin1} (Pd$_{0.99}$Fe$_{0.01}$, Pd$_{0.9}$Ni$_{0.1}$ and Cu$_{0.4}$Ni$_{0.6}$, respectively).

\begin{table}
\caption{Transport properties of single, bi and trilayer SFS JJs. All the parameters have been determined at 0.3 K, while $T_c$ is defined as the temperature at which $I_c$ = 0, i. e. the JJs present a linear IV characteristic.}
\label{tab:1}
\begin{tabular}{llllllll}
\hline
      Type of junction &  $I_c$ ($\mu$A)   &   $J_c$ (kA/cm$^2$) & $I_c R_N$ ($\mu$V) & $T_c$ (K)               \\ \hline
    Single layer (Co)       &      410           &                80            & 5.1            &          3.0          \\     
    Single layer (Co)       &      210           &                40            & 2.0            &          4.0          \\ 
    Single layer (Co)       &      170           &                38            & 3.2            &          2.5          \\ \hline
    Bilayer  (Ho/Co)         &      210           &                47           &  130             &          4.0           \\ 
    Bilayer  (Ho/Co)         &       40            &                 8            &  25             &          2.5            \\ 
    Bilayer  (Ho/Co)         &       750          &               100          &  70             &          5.0            \\ \hline
    Trilayer (Ho/Co/Ho)    &       80            &                16           &   430           &          1.5            \\
    Trilayer (Ho/Co/Ho)    &       20            &                 4            &  210           &           3.0           \\ \hline
    \end{tabular}
\end{table}

The origin of hysteresis in fully metallic SFS junctions with strong F layers requires careful consideration to distinguish the relative contributions from capacitive and self-heating effects. Figure \ref{fig:1}d shows the $I_c R_N$ product of the junctions as a function of the switching voltage $V_{sw}$ (left axis, full symbols). For the same junctions, the hysteresis is shown by the open symbols (right axis). Here, the $I_c R_N$ product and the hysteresis ($H_y$) scale as a function of $V_{sw}$. Additionally, $I_c R_N$, $V_{sw}$ and $H_y$ increases with the number of F layers in the junction. $I_c R_N$ product ranges from \cite{blamire2012} a few $\mu$V for single layer JJs up to a few hundreds of $\mu$V for bilayer and trilayer JJs, with a maximum value of about 500 $\mu$V and the corresponding hysteresis of about 70\%. At the same time, the Joule power deposited in the weak link, calculated as $I_c V_{sw}$, ranges between a fraction of a nW up to a few tens of nW. It generally increases with the number of F layers, but the trend is not clear since most of the trilayer junctions are characterized by lower values of Joule power with respect to bilayer JJs. Such high values of Joule power may suggest the presence of electron overheating in the weak link \cite{pekola}. Therefore, the scaling of the $I_c R_N$ product and of $H_y$ as a function of $V_{sw}$ does not solve the ambiguity between self-heating processes and effective capacitance contributions. More insights on the electrodynamics come from the measurements and the analysis of the temperature dependence of the critical current and of the switching current distributions, reported in the next two Sections.

\section{Temperature dependence of the critical current}
\label{energy}

\begin{figure}[htb!]
\begin{center}
\includegraphics[width=0.8 \linewidth]{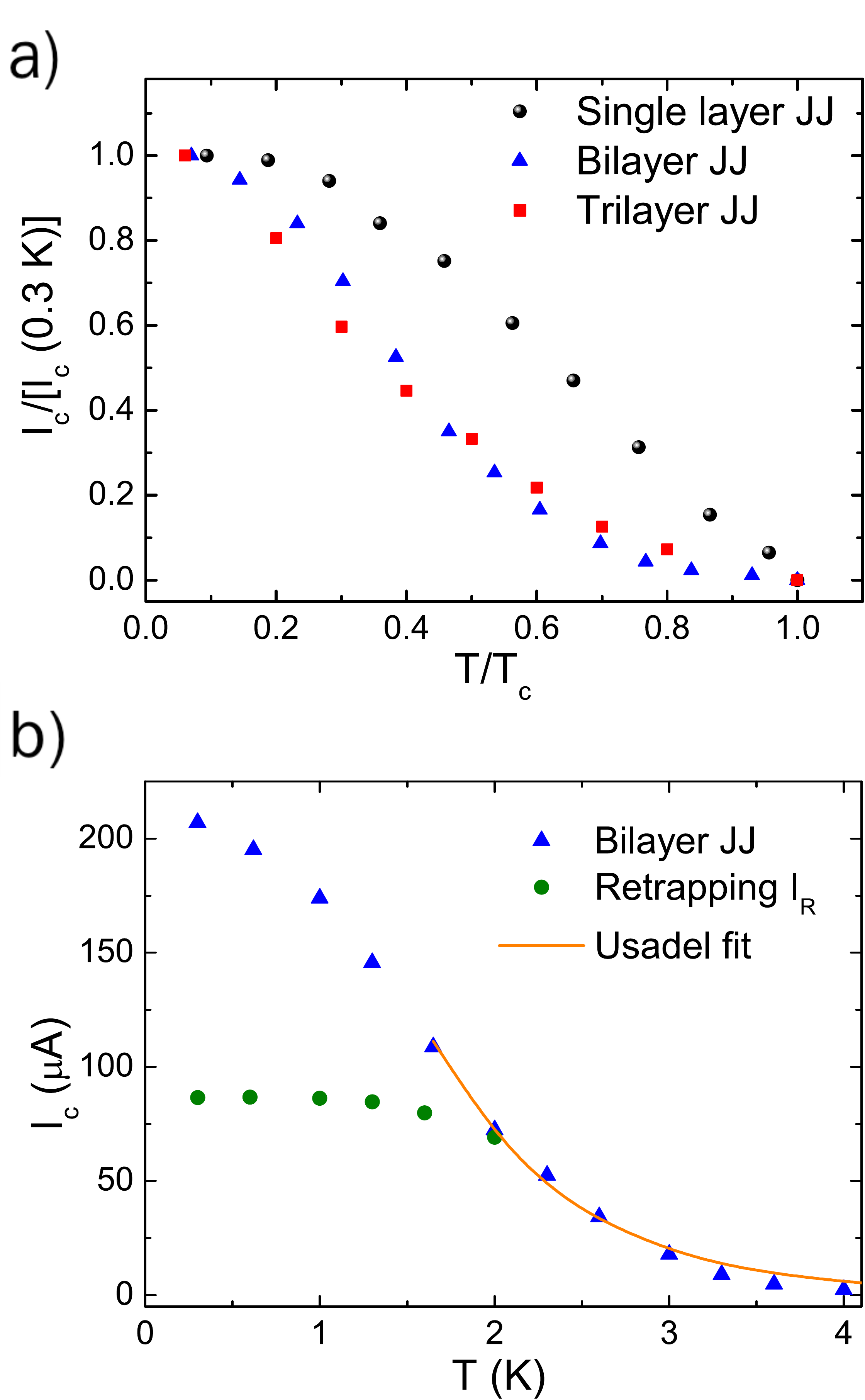}
\caption{a) Temperature dependence of the critical current $I_c$ of single layer (black dots), bilayer (blue triangles) and trilayer (red squares) JJs, respectively. $I_c$ is normalized to the value measured at 0.3 K, while $T$ is normalized to $T_c$. Once normalized, the curves for bilayer and trilayer JJs overlap and show a characteristic exponential behavior for $T>$ 0.3 $T_c$, while a linear trend at high temperatures has been observed in single layer JJs. In panel b) the $I_c$ vs $T$ measurements are reported for the bilayer JJ (blue triangles), without normalization. The orange curve is the Usadel fit for $T \geq$ 1.5 K and provides an estimation of the Thouless energy of about 25 $\mu$eV (see the text). The retrapping current $I_R$ is shown by the green circles. For $T\geq$ 2 K, $I_R = I_c$ and no hysteresis is present in the IV characteristics.}
\label{thouless}
\end{center}
\end{figure}

In long SNS junctions the energy scale for the proximity effect is given by the Thouless energy. In contrast to the energy gap $\Delta$, which is set by the interactions in the superconducting electrodes, the energy scale $E_{th}$ is a single-electron quantity and is related to the diffusion rate across the sample for a single electron. For SFS junctions where triplet supercurrents dominate the transport, a long junction means that both the conditions $L > \xi_F$ and $L> l_e$ are satisfied, as discussed in Sections \ref{setup} and \ref{IV}. A clear fingerprint of long junction regime is given by the temperature behavior of the critical current. In Fig. \ref{thouless}a the $I_c$ vs $T$ behavior is reported for single layer (black dots), bilayer (red triangles) and trilayer (blue squares) JJs. In this plot $I_c$ is normalized to the value measured at 0.3 K and the temperature is normalized to $T_c$, estimated as the temperature at which the IV characteristic becomes a linear curve. The measurements shown in Fig. \ref{thouless}a are representative of a general behavior: single Co layer JJs show an almost linear behavior at high temperatures and tend to saturate below 0.2 $T_c$, while the $I_c$ vs $T$ dependences for bilayer and trilayer junctions are quite similar. They have a characteristic exponential behavior with an upward curvature for $T$ $>0.3$ $T_c$, typical of long SNS JJs. The long-junction behavior is expected since the high values of the characteristic voltage in bilayer and trilayer junctions show that the supercurrent is mediated by triplet Cooper pairs. According to Refs. \onlinecite{zaikin2001,angers2008}, in the limit $\Delta >> E_{th}$ and for temperatures such that $k_B T >$ 5 $E_{th}$, the $I_c$ vs $T$ dependence of long JJs is given by the following Usadel equation:
\begin{equation}
	I_c=\frac{64 \pi k_B T}{eR_N} \sqrt{\frac{2\pi k_B T}{E_{th}}} \frac{\Delta ^2 \exp [-\sqrt{\frac{2\pi k_B T}{E_{th}}}]}{ [ \omega_0 + \Omega_0 + \sqrt{2(\Omega_0 ^2 + \omega_0 \Omega_0)} ]^2 }
      \label{eqzaikin}
\end{equation}
where $\omega_0= \pi k_B T$ is the zero order Matsubara frequency and $\Omega_0 = \sqrt {\Delta^2 + \omega_0 ^2}$.

In Fig. \ref{thouless}b the $I_c$ vs $T$ data are reported for a bilayer junction, along with the temperature behavior of the retrapping current, which indicates that hysteresis is present below about 2 K and that $I_R$ levels off below 1 K, while $I_c$ still increases at low temperatures. A fit of the high temperature data, above 1.5 K, by using the Usadel equation is in very good agreement with the experimental behavior and allows us to estimate the Thouless energy, which is about 25 $\mu$eV for the junction reported in Fig. \ref{thouless}b. To perform the fit, the superconducting gap has been determined from $\Delta= 1.76 k_B T_c^{Nb} \simeq$ 1.2 meV, where $T_c^{Nb}$ is the critical temperature of the superconducting electrodes. Moreover, a better fit has been obtained when the BCS temperature dependence of the gap has been taken into account, as in Ref. \onlinecite{zaikin2001}. Finally, both conditions $\Delta \gg E_{th}$ and $k_B T>$ 5 $E_{th}$ are self-consistently satisfied (the equivalent temperature $T_{th}$ is about 280 mK). The same analysis has been performed on other bilayer and trilayer junctions and the Thouless energy falls in the range between 20 and 40 $\mu$eV.

\section{Phase dynamics}
\label{phase}

Measuring the SCDs as a function of the temperature is a powerful tool to investigate the phase dynamics of JJs \cite{fulton,martinis1987,voss}. Different dissipation processes, which are not accessible through the analysis of the IV characteristics, can be detected by the thermal dependences of the switching histograms \cite{martinis1987,vion,noiLTP,breakdown}. Phase diffusion phenomena in the moderately damped regime \cite{kautz,krasnov2005,noiLTP,krasnov2007,prbLuigi,kivioja2005,mannik2005,luigi2,daniela} or local heating events induced by non-equilibrium processes in high $J_c$ junctions \cite{breakdown} have distinctive fingerprints given by the thermal behaviors of the SCDs and their first three central momenta: the mean switching current $I_m$, the standard deviation $\sigma$ and the skewness $\gamma$, respectively.

\begin{figure}[htb!]
\includegraphics[width=0.78 \linewidth]{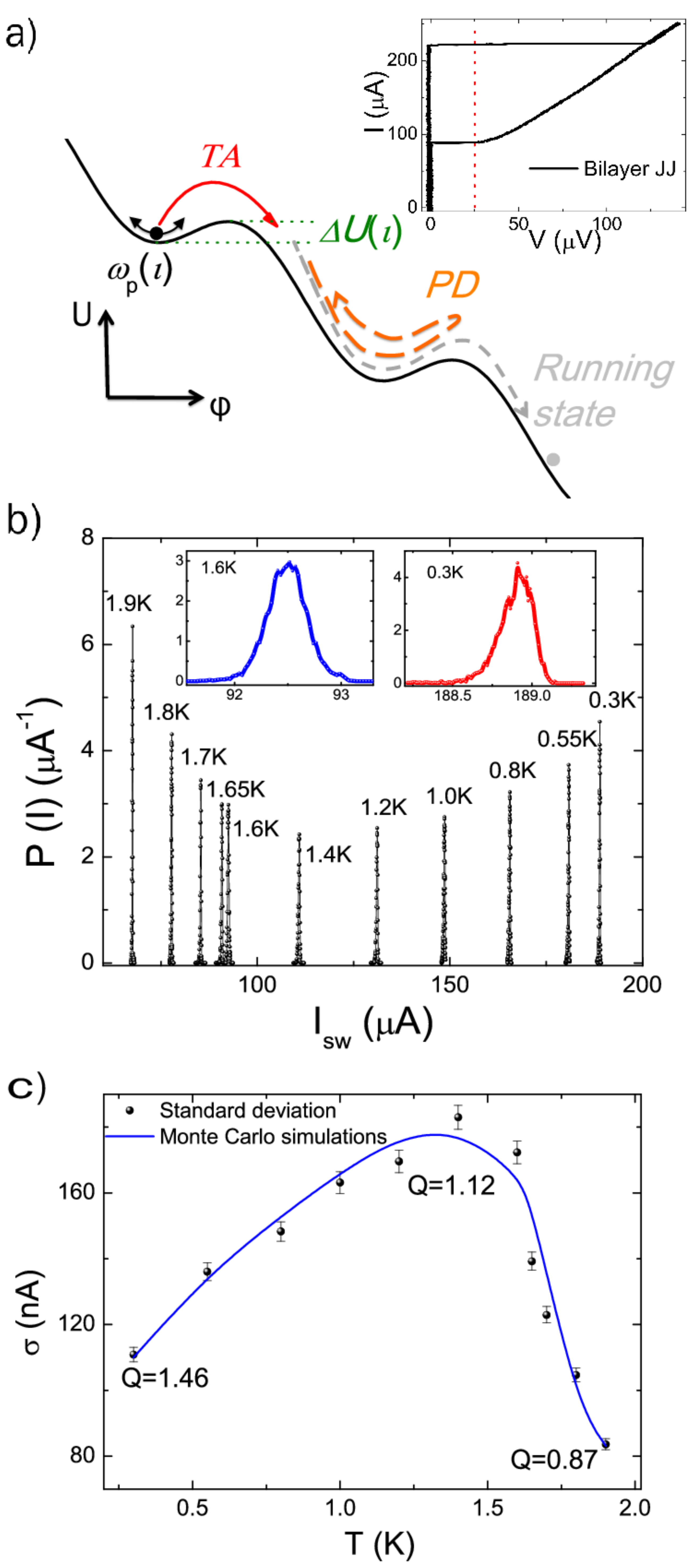}
\caption{a) Dynamics of a phase particle in a tilted washboard potential for $i$ slightly less than 1. Thermal activation (TA) above the barrier $\Delta U (i)$ (green dotted lines), retrapping processes in the phase diffusion (PD) regime and running motion along the potential are qualitatively sketched by the red arrow, orange dashed line and grey dashed line, respectively. The inset shows the IV curve of a bilayer JJ, with the threshold voltage (red dashed line) of about 20 $\mu$V for the switching measurements. b) Measurements of SCDs as a function of the temperature for a bilayer JJ. The insets show the zoom of the SCDs measured at 0.3 K (red dots, asymmetric distribution in the TA regime) and at 1.6 K (blue dots, symmetric distribution in the PD regime). The lines are guides for the eye. c) Temperature behavior of the standard deviation $\sigma$ (black dots), extracted from the SCDs reported in panel b). The blue line is the fit obtained by Monte Carlo simulation of the phase dynamics, with a quality factor Q = 1.46 at 0.3 K. At $T^*$ = 1.4 K Q = 1.12 and at 1.9 K Q = 0.87. Above 2 K the IV curves are non-hysteretic, see Fig. \ref{thouless}b.}
\label{distribution}
\end{figure}

According to the RCSJ model \cite{barone,likharev}, the phase dynamics of a JJ is analogue to the motion of a particle in a tilted washboard potential $U(\varphi)= -E_{J}(\cos\!\varphi + i \varphi)$ (see Fig. \ref{distribution}a). Here, $E_{J}=I_{c0}\phi_0/2\pi$ is the Josephson energy, $\phi_{0}=h/2e$ is the quantum flux, $I_{c0}$ is the critical current in absence of thermal fluctuations and $i=I/I_{c0}$ is the normalized bias current, which determines the tilt of the potential. The motion of the particle is subject to damping given by $1/Q$, where $Q$ is the quality factor defined in Section \ref{IV}. When the bias current is ramped from $i = 0$ to $i <1$, the junction is in the zero voltage state in absence of thermal and quantum fluctuations, and the phase particle is confined to a potential well, where it oscillates at the plasma frequency $\omega_p (i) = \omega_p (1-i^2)^{1/4}$. At finite temperature the junction may switch into the finite voltage state for a bias current $i <1$. Due to thermal fluctuations, the phase particle can overcome the potential barrier $\Delta U (i) = 4\sqrt{2}/3 \cdot E_J (1- i)^{3/2}$. This regime is known as Thermal Activation (TA) and the escape rate is determined by \cite{kramers} $ \Gamma_{t}= a_{t} \frac{\omega_{p} (i)}{2\pi} \exp\left(-\frac{\Delta U (i)}{k_{B}T}\right) $, where the thermal prefactor is \cite{but83} $a_{t}=4 \cdot \left[\left(1+Qk_{B}T/1.8\Delta U \right)^{1/2}+1\right]^{-2}$. TA is the main escape process at high temperatures $k_BT \gg \hbar \omega_p$ and is qualitatively sketched in Fig. \ref{distribution}a by the red arrow.

In the underdamped regime, the escape from the metastable state corresponds to the appearance of a finite voltage across the junction. As it is shown by the grey dashed line in Fig. \ref{distribution}a, the escaped particle gains sufficient energy to roll down the potential in the so-called running state. In case of moderately damped JJs ($Q\geq$ 1) the dynamics is different, since escape due to thermal hopping does not lead to runway down the tilted potential \cite{kautz}. After the escape event, the particle can be retrapped in one of the following minima of the potential, as sketched by the orange dashed line in Fig. \ref{distribution}a. Multiple escape and retrapping processes induce a diffusive motion of the phase particle along the washboard potential before switching to the resistive state, and this regime is known as phase diffusion (PD) \cite{kautz,krasnov2005,noiLTP,krasnov2007,prbLuigi,kivioja2005,mannik2005,luigi2}. 

The experimental switching probability density $P(I)$ is related to the escape rate $\Gamma(I)$ through the following equation \cite{fulton}:

\begin{equation}
	P(I)=\frac{\Gamma (I)}{\Delta I/\Delta t} \exp \left[- \int_{0}^{I} {\frac{\Gamma (I')}{\Delta I'/\Delta t}dI'} \right]
      \label{fulton}
\end{equation}

Fig. \ref{distribution}b shows a set of SCDs as a function of the temperature, from 0.3 K up to 1.9 K, for the same bilayer SFS JJ reported in Fig. \ref{thouless}b, with a zoom on the histograms measured at 0.3 K (red curve) and at 1.6 K (blue curve) in the inset. Due to the strong temperature dependence of the critical current and to the very low values of the ratio $\sigma/I_{m} \simeq 10^{-3}$, the SCDs cover a very large range of switching currents and are quite narrow. In this temperature range, the behavior of the SCDs is typical of moderately damped JJs \cite{kautz,krasnov2005,noiLTP,krasnov2007,prbLuigi,kivioja2005,mannik2005,luigi2,daniela}: the standard deviation $\sigma$, shown in Fig. \ref{distribution}c, increases in the temperature range from 0.3 K up to 1.4 K, then it starts to collapse indicating the transition to the PD regime. We identify $T^*$ = 1.4 K as the transition temperature between the TA and the PD regime. The insets of Fig. \ref{distribution}b also compare two SCDs measured at 0.3 K and at 1.6 K. In the former case, the SCD is asymmetric with the characteristic tail on the ascending side of the hystogram, typical of the TA regime, while in the latter case the SCD is more symmetric due to the onset of retrapping processes in the PD regime \cite{krasnov2005,noiLTP,krasnov2007,prbLuigi,kivioja2005,mannik2005,luigi2,daniela}. The experimental results on other bilayer and trilayer JJs are quite similar and with almost the same temperature behavior of $\sigma$. For single layer JJs, we could not measure the SCDs due to the very low values of the switching voltage, of the order of a few $\mu$V. A threshold voltage of about 20 $\mu$V, as the one shown as red dashed line in the inset of Fig. \ref{distribution}a, is necessary to distinguish the switching events from spurious noise. 

By fitting the switching probability density $P(I)$ in the TA regime using Eq. (\ref{fulton}), the critical current in absence of thermal fluctuations $I_{c0}$ can be estimated at each temperature. More importantly, the temperature behavior of $\sigma$ can be reproduced through Monte Carlo simulations of the phase dynamics: the fitting parameter is the quality factor $Q$ which regulates the collapse of $\sigma$ and the transition temperature $T^*$ \cite{prbLuigi, luigi2}. In the simulations, the phase difference $\varphi(t)$ is a solution of the following Langevin differential equation:

\begin{equation}
	\varphi_{tt} + \varphi_t /Q + i + i_N = 0
      \label{giacomo}
\end{equation}

Times $t$ are normalized to $\omega_p^{-1}$, and $i_N$ is a Gaussian correlated thermal noise current, such that:

\begin{equation}
	\left< i_N(t) \right>=0; \hspace{5 mm} \left< i_N (t) i_N (t')\right>= \sqrt{k_B T/Q E_J} \delta(t-t').
      \label{giacomo2}
\end{equation}

Stochastic dynamics are simulated by integrating the above Langevin equation by a Bulirsh-Stoer integrator using as noise generator the cernlib routine RANLUX \cite{ranlux}. Simulations have been carried out for different temperatures and dissipation values. The multiplicity of switching modes between the running and the trapped states raises a problem of how to define an escape event. In our simulations, the escape event is declared when the phase particle spends in the running state more than 50\% of the observation time. Typical runs for simulations of Eq. (\ref{giacomo}) last from 4$\cdot$ 10$^6$ to 6$\cdot$ 10$^6$ normalized time units, that is, 6$\cdot$ 10$^5 $ to 9$\cdot$ 10$^5$ plasma periods. Observation time for each point generated in the IV characteristics is 2$\cdot$ 10$^4$ time units, which is a long enough time to ensure that the average time spent in running/zero voltage state does not vary as a function of the observation time.

To obtain the SCDs we have simulated a number of escape events between 3000 and 5000, which is similar to the number of counts experimentally collected. More details on Monte Carlo simulations can be found elsewhere \cite{prbLuigi, luigi2}. In contrast to previous works, in the simulations the quality factor here is temperature dependent, reflecting the strong temperature dependence of $I_c$. The best fit is shown as the blue curve in Fig. \ref{distribution}c with $Q = 1.46$ at 0.3 K, $Q = 1.12$ at $T^*$ and $Q =0.87$ at 1.9 K. Above this temperature hysteresis is almost zero. According to numerical simulations reported in Ref. \onlinecite{kautz}, the IV characteristic is hysteretic for $Q \geq$ 0.84, providing a further proof of the consistency of the Monte Carlo fit. Finally, it is worth mentioning that a variation of 0.01 in $Q$ provides different $\sigma$ vs $T$ curves, with less agreement with the experimental data.

Therefore, measurements of SCDs point to a Josephson dynamics with a defined $Q$ factor slightly larger than 1. Such a large value of $Q$ for a SF(N)S JJ can be explained by considering that, while geometric capacitance in these types of structures can be very small, the presence of Andreev bound states in the N (F) layer yields additional capacitance-like contributions, which can dominate over the geometric capacitance \cite{zaikin2001,zaikin2010,skvortsov2015}. The intrinsic capacitance can generally be estimated as \cite{skvortsov2015} $C_i = a_c \hbar \cdot (R_N E_g)^{-1}$, where $E_g$ is the proximity minigap and the $a_c$ coefficient is of the order of 0.9 for long junctions \cite{skvortsov2015}. In that limit, $E_g$ = 3.12 $E_{th}$, therefore from the Thouless energy estimated by the $I_c$ vs $T$ fit reported in Section \ref{energy}, we obtain $E_g \simeq$ 77 $\mu$eV and $C_i \simeq$ 8 pF for the bilayer junction reported in Figs. \ref{thouless} and \ref{distribution}.

It is well-known that the effective damping in tunnel SIS junctions is typically dominated by the high frequency impedance of the circuitry \cite {martinis1987}, which in general is of the order of 50-100 $\Omega$. In SNS junctions, where the $R_N$ resistance is of the order of a few $\Omega$ or less, the shunting by the high frequency impedance of the circuitry is avoided, thus the effective damping is dominated by $R_N$ itself. Indeed, by considering the values of $Q$ from the $\sigma$ vs $T$ fit and of $C_i$, we obtain an effective resistance of about 0.7 $\Omega$, which is very close to $R_N$ ($\simeq$ 0.6 $\Omega$). Capacitance renormalization also affects the plasma frequency $\omega_p$: by considering the capacitance entirely arising from geometric considerations of the order of 1 fF, one would obtain $\omega_p \approx$ 4 THz, which is an unrealistic value for low critical temperature JJs. However, by considering the role of the intrinsic capacitance, $\omega_p$ is of the order of 45 GHz at zero bias, 25 GHz for $I = I_m$, when the switching to the resistive state occurs. These values are more realistic and routinely observed for underdamped and moderately damped JJs \cite{mqt,fulton,voss,martinis1987,prbLuigi}.

\section{Discussion and concluding remarks}
\label{disc}

The combined analysis of SCDs and IV characteristics as a function of the temperature provides a self-consistent picture for the electrodynamics of bi and trilayer SFS junctions with strong ferromagnetic interlayers. An unconventional behavior has been observed in these JJs where the supercurrent transport is strongly dominated by triplets. This anomalous behavior manifests in the temperature dependence of the critical current, where the bilayer and trilayer junctions show a long-junction behavior in contrast to junctions containing a single ferromagnetic layer, thus allowing us to estimate the Thouless energy, while the temperature behavior of the switching current distributions clearly points to a moderately damped dynamics. The relevant hysteresis in the IV characteristics, which builds up when going down to temperatures lower than 2 K, can be explained  by considering a renormalization of the RCSJ parameters: the dominating capacitance term is the intrinsic capacitance due to the Andreev levels in N (F) layer, while the effective resistance is close to the normal state resistance. The resulting quality factor is slightly larger than 1, in agreement with the moderately damped dynamics which results from the SCDs measurements. According to classical proximity models \cite{zaikin2001,zaikin2010,skvortsov2015}, the intrinsic junction capacitance, due to the dynamics of the Andreev bound states, can be explained only assuming that bi and trilayer junctions behave as effective SNS JJs, with an equal-spin triplet component with a decay length comparable to $\xi_N$ \cite{efetov_rew}.

At the same time, the switching dynamics related to local heating events and non-equilibrium phenomena, as those recognized in Ref. \onlinecite{breakdown}, have not been observed, thus confirming a proper RCSJ dynamics of the measured junctions. This does not exclude possible electron overheating in the N (F) layer after the switching, which affects or produces a finite hysteresis in the IV curves. Nevertheless, the good agreement between data and Monte Carlo simulations with $Q \leq$ 1 at high temperatures, when hysteresis is going to disappear, suggests that, if present, electron overheating should play a minor role.

These results fall under the general framework of JJs with intermediate values of the critical current density $J_c$ and large interface transparencies, and represent the first electrodynamic characterization of fully metallic SFS junctions containing strong ferromagnets. In the last few years, composite barriers made up of strong (Co) and weak (Ho) ferromagnet have represented the model system exhibiting triplet superconductivity, and can be potentially employed as cryogenic memories, $\pi$-phase shifters, spintronic elements in more complex circuits. In this work, we have provided clear fingerprints of the phase dynamics with the detailed set of electrodynamic parameters of such junctions, which show strong evidence of long range triplet Cooper pairs, thus stimulating further studies for a possible different way to detect triplet supercurrents.

 \begin{acknowledgments}
 
Enlightening discussions with A. Golubov, P. Lucignano and G. P. Pepe are gratefully acknowledged. DM, RC, FT would like to thank NANOCOHYBRI project (Cost Action CA 16218). NB acknowledges funding from the British Council through UKIERI programme and Loughborough University. MGB acknowledges funding from EPSRC Programme Grant EP/N017242/1.

\end{acknowledgments}

\end{document}